\documentstyle[11pt,aaspp]{article}

\def\spose#1{\hbox to 0pt{#1\hss}}
     
\def\ltsim{\mathrel{\spose{\lower 3pt\hbox{$\mathchar"218$}}
     \raise 2.0pt\hbox{$\mathchar"13C$}}}
\def\gtsim{\mathrel{\spose{\lower 3pt\hbox{$\mathchar"218$}}
     \raise 2.0pt\hbox{$\mathchar"13E$}}}
\def\etal{{\it et al.}}
     
\def\eg{{\it eg.\ }}     

\begin{document}

\title{The Mass and Extent of the Galactic Halo}

\author{Dennis Zaritsky}
\affil{UCO/Lick Observatory, Univ. of Calif. at Santa Cruz}

\begin{abstract}
We review the various techniques
used to constrain the Galactic halo 
mass profile for radii $\gtsim $ 20 kpc with particular
emphasis on identifying a self-consistent halo model and resolving
the apparent discrepancies present in the literature on this subject. 
We collate published results and demonstrate that they are 
all consistent with a Galactic halo that is nearly isothermal with
a characteristic velocity of 180 to 220 km sec$^{-1}$ and an 
extent $\gtsim$ 200 kpc. 
\end{abstract}

% Keywords should be included, but they are not printed in the hardcopy.
% An attempt to select keywords following the U of Chicago Press subject
% headings (available on-line at http://www.noao.edu/apj/keywords96.html)
% would be greatly appreciated!

\section{Introduction}
Two fundamental parameters that characterize the 
Galactic halo are its mass and extent. Our understanding
of these two quantities 
frames our subsequent investigations of the halo and provides basic
constraints on models of Galactic formation and evolution. For example, if
the halo is unequivocally vastly larger and more massive
than the luminous component of the Galaxy, then the search for dark matter
at large radius becomes a key research endeavor. 
This review describes the current understanding of
the mass distribution of our Galaxy at large ($>$ 20 kpc)
radii with particular emphasis at reconciling a single model with
all of the available data.

\section{Preliminaries}

What is the Galactic halo? 
Some investigators use the term halo when describing material 
just outside of 
the disk (\eg 1 to 2 kpc above the disk), 
while others use it only when describing the material well 
beyond radii of several tens of 
kpc. For some investigators, the halo is the spherical stellar
component 
of the galaxy, for others it is only the dark matter component.
No particular definition is superior, but it is essential in any
discussion to describe one's adopted definition. Here, we
use the term halo to describe the mass distribution external to $\sim
20$ kpc. 

The concept of ``the mass of the Galactic halo" is ill-defined. If 
models (cf. Navarro,
Frenk \& White 1996) of galactic halos are even remotely correct, then halos 
do not have sharp boundaries and galaxies are sufficiently close to each 
other that their halos should overlap. Because halos are not discrete,
finite objects, ``total'' quantities (such as total mass or extent)
are unmeasurable. Galaxy masses are commonly quoted in the
literature, but generally the authors implicitly define that mass 
to be the mass within the radius that they probed.
Again, great care must be taken to explicitly
state one's definition and to be consistent when comparing results
from various studies.

Is there an acceptable practical, working definition of a halo?
One potential definition is that the halo is the volume enclosing 
all of the mass that
has already decoupled from the Hubble expansion.
Although this definition is
well-defined theoretically, it is difficult to implement in
practice because the current turnaround radius is unobserved and because
galaxies are not isolated. For example, although M 31 has decoupled
from Hubble flow and is falling toward the Milky Way, we would
not consider M 31 to be part of the Galactic halo. Other
theoretical definitions such as the mass enclosed within the virial radius,
the mass that is gravitationally bound, or the mass within a fixed
density contrast relative to the universal mean density, 
are equally problematic --- especially when applied to real galaxies.
The only viable solution to this problem is to avoid defining
the halo as a discrete entity. Instead, we must focus our discussion
on the mass profile or on the mass within a selected, fixed radius.

The emphasis on the mass profile or the enclosed mass within radius, $R$,
raises a second issue. The test particles utilized
to measure the mass (e.g. H I atoms, globular clusters, satellite galaxies,
or nearby galaxies) {\bf must} span the radial range over which the
enclosed mass is being quoted. For example, one should never rely on
the Galactic disk rotation curve, which at best extends to $\sim$
20 kpc (Fich \& Tremaine 1991), to infer an enclosed mass at $R > 20$
kpc.  Such an
extrapolation requires that one assume a halo profile and extent
beyond 20 kpc
--- assumptions that turn the analysis into a self-fulfilling prophecy.
Such studies (cf. Honma \& Ken-ya 1998) 
typically note a slight decrease or increase
in the outermost rotation curve that, when extrapolated, leads
to large apparent discrepancies with, for example, the halo mass
inferred from Galactic satellite dynamics. {\bf Most
of the apparent discrepancies between different studies are the result
of comparing masses derived for different effective enclosed radii
or from the extrapolation of results beyond the radii that were
directly probed.} For illustration, Figure 1 shows the relative
sizes (in projection) for the most common regimes discussed in the 
literature (disk, interior to the LMC, interior to the outermost 
satellites, and theoretical halo). Notice that
the disk and even the LMC at its current position, probe a
relatively small fraction of the what can be considered to be
the Galactic halo.

\begin{figure}
\plotone{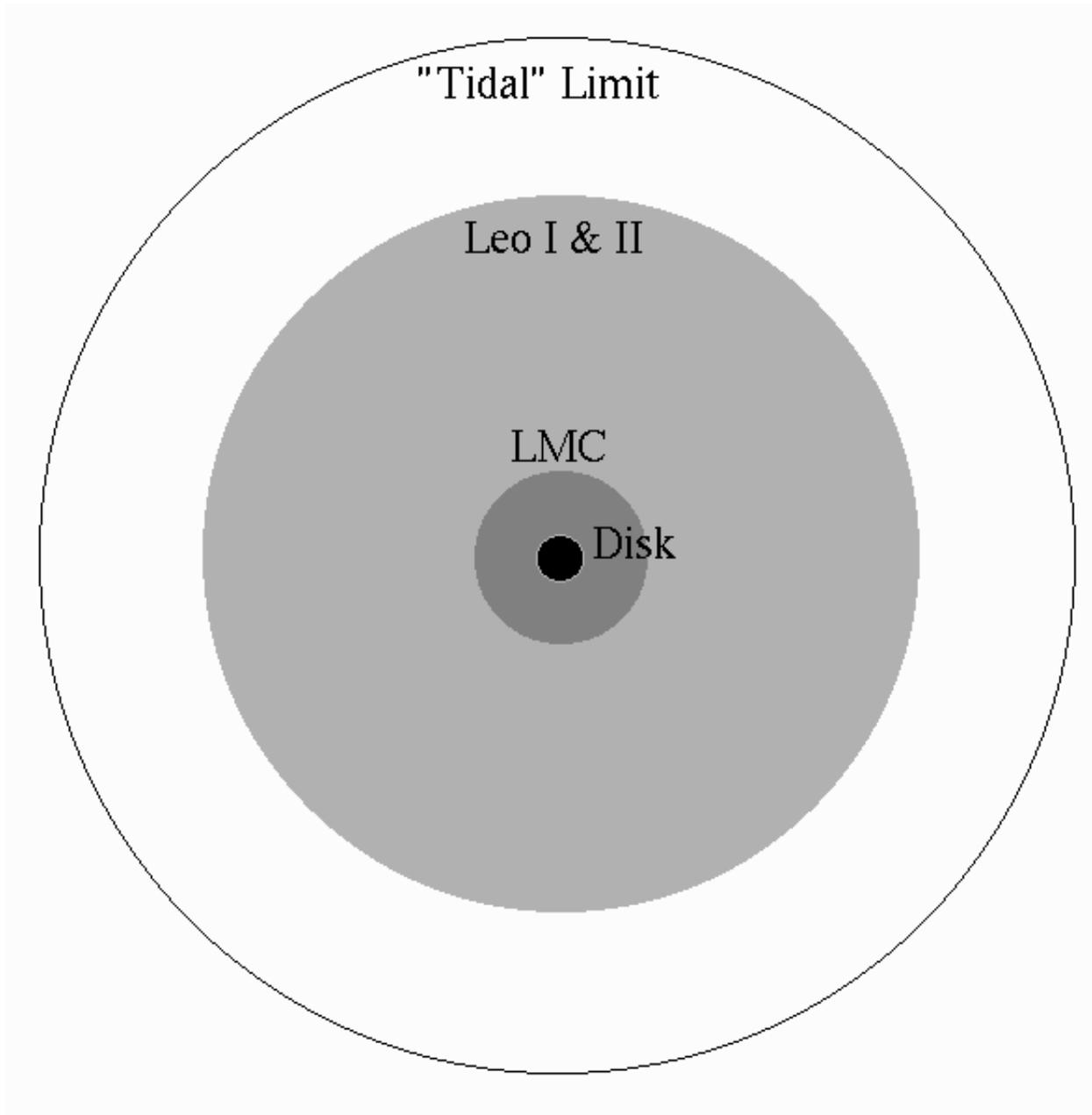}
\caption{A comparison of the range over which various probes 
sample the Galactic halo. The plotted areas are scaled relative to
the projected area of the halo enclosed by the probe ---  
a plot of the relative volumes probed would
show an even larger difference between the inner and outer regions.}
\end{figure}

The final issue to be discussed involves the rather unsavory, but
widespread practice of selective citation. A proper halo model
must be in accordance with all of the available data (to within the 
associated uncertainties) ---
although one does not need to accept all of the available conclusions! 
For example, it is inappropriate to cite only the measures of the
Galactic mass inside of 50 kpc (e.g. rotation curve and the kinematics of
the inner globular clusters) as a measure of the mass of the Galactic halo.
Not only do they not probe beyond 50 kpc (and are therefore almost
entirely insensitive to whatever mass might be present beyond 50 kpc),
but if the Galaxy had such a small total mass, one would not
be able to account for the dynamics
of the outer satellite systems and the dynamics of the Local Group.
Therefore, the omission of the probes at larger
radii does not simply lead to  ``a conservative mass estimate'' but
instead leads to a result that is inconsistent with existing data.
Every data set and analysis approach 
includes some assumption(s) that if violated by nature
would result in a serious under or overestimate of the Galactic mass.
Because the uncertainties are dominated by systematic errors,
the use of {\it all} of the data strengthens the conclusions
by more than the standard $1/\sqrt{N}$ intuitive understanding.

With these caveats in mind, we proceed to briefly discuss the principal
methods for the determination of the mass profile of our Galaxy for
$R > $ 20 kpc.

\section{The Methods}

\subsection{The Rotation Curve}

As mentioned previously, the 
rotation curve of our Galaxy does not provide strong leverage on
the mass profile for $R >$ 20 kpc, but it does provide a key anchor
for the mass profile at small radius. Models used to explain the
dynamics of the outer halo cannot predict velocities at small
radii that are inconsistent with the precise measures of the disk
rotation. For example, although
it is plausible that the outer halo systems
might be well fit by an isothermal halo with $v_c = 300$ km
sec$^{-1}$, such a model can be rejected because 
it predicts that the rotation curve of the galaxy at all $R$
is 300 km sec$^{-1}$. 

The measurement of the outer rotation curve of our galaxy is
complicated by our position within the disk. The exact values
of $v_c(15$ kpc) has oscillated mostly between 180 km sec$^{-1}$
and 220 km sec$^{-1}$ (Fich \& Tremaine 1991) and the
uncertainties are large as $R \rightarrow 20$ kpc. 
Our final model should include the full range of
these possibilities, but should not over interpret a rise
or decline in the rotation curve at large radii. 
For an isothermal sphere model, the implied
mass ratio between the two extremes is only $(180/220)^2
= 0.67$, which in this endeavor is not considered to be a serious
discrepancy. 

A strikingly different conclusion is reached if one interprets the
decrease in the rotation curve from 220 km sec$^{-1}$ at 8 kpc to 180
km sec$^{-1}$
at 15 kpc as due to a centrally concentrated mass distribution. 
This velocity drop is nearly consistent with a Keplerian fall-off
and would imply a central mass of about $1.1 \times 10^{11} M_\odot$.
The rotation speed values recently derived by Merrifield
and Olling (1998) of 166 km sec$^{-1}$
at $R = 20$ imply, for an {\it assumed} Keplerian rotation curve,
a Galactic mass of $1.2 \times 10^{11} M_\odot$.
These values for the Galactic mass 
are a factor of ten smaller than the mass inferred
at 200 kpc from the dynamics of satellite galaxies. However, this
discrepancy is completely fictitious and entirely produced by
the extrapolation of the Keplerian
rotation curve to large radii (for which the rotation curve provides
no evidence). If instead we ask what rotation velocity
is implied at 15 or 20 kpc for an isothermal halo normalized
to have an enclosed mass 
of 1.2$ \times 10^{12} M_\odot$ at $R=200$ (a value shown later to be
implied by the distant Galactic satellite galaxies), 
the characteristic velocity is
$v_c$ = 165 km sec$^{-1}$ --- actually lower
than that measured by the rotation curve at $R =15$ kpc and in 
agreement with Merrifield and Olling's value at $R=20$ kpc! 
Therefore, the observed
rotation curve places no significant constraints on models of the
Galactic halo that include a massive isothermal-like component that
is consistent with the dynamics of the outer halo (obviously,
the rotation curve is 
also consistent with a halo that truncates shortly outside
$R = 20$ kpc). This exercise demonstrates why the extrapolation of 
data from one regime to larger radii is perilous.

\subsection{Stellar Escape Speed}

Stars observed locally are presumably bound to the Galaxy.
Therefore, the fastest moving stars
place a lower limit on the local escape velocity (cf. Carney \&
Latham 1987). An analysis of these
stars suggests that the local escape
speed is between 450 and 650 km sec$^{-1}$ (Leonard \& Tremaine
1990). Converting the observed velocities into an estimate of the
escape speed requires a model for the Galactic potential and 
the stellar velocity
distribution. While the fastest observed star provides a well-defined
lower mass limit, the derivation of an upper limit is highly model 
dependent. Applying a Jaffe (1983) model (which provides a flat
rotation curve at small radii and a sharp, $\rho \propto r^{-4}$,
cutoff at large radii, and an adopted circular velocity of 220 km
sec$^{-1}$ at small radii, the lower limit on the escape velocity
provides a lower mass limit at 200 kpc of $3.8 \times 10^{11} M_\odot$
(Kochanek 1996). This is an interesting lower mass limit because the
inferred scale length, $r_j$, of the Jaffe model is 44 kpc and is roughly
equivalent with the current position of the Large Magellanic Cloud (LMC)
--- thereby suggesting that any dark halo that exists must extend at
least to the current position of the LMC. If this minimal halo
is characterized as an isothermal sphere, then 
$v_c \gtsim 180$ km sec$^{-1}$.

\subsection{Distant Satellite Galaxies and Globular Clusters}

The most distant probes, that are still plausibly within the Galactic
halo, globular clusters and satellite
galaxies at $R > $ 50 kpc. These objects
provide the best opportunity to measure the 
outer halo mass profile,
but suffer from other difficulties. There are few such
objects ($\sim 15$ at $R \ge 50$ kpc), typically only their radial
velocity is known, their velocity ellipsoid is unknown, the outermost ones
may not be gravitationally bound to the Galaxy, and
they have sufficiently large orbits that in a Hubble time they have
completed only 1 or 2 orbits. Some of these problems are being
addressed, for example proper motions have been measured for the 
nearest satellites (Majewski \& Cudworth 1993).

The properties of the satellite sample can be analyzed to provide
constraints on the mass and extent of the halo. The simplest approach
is to apply the escape velocity argument to these objects. Assuming
a point mass potential ({\it i.e.} all of the mass is within the current position of
the satellite), the radial velocity of the Leo I satellite (177 km
sec$^{-1}$ in the local standard of rest) and the distance to Leo
I (230 kpc from the Galactic center), we calculate that $M_{MW} > 8 \times 10 
^{11} M_\odot$ (for a listing of satellite data and references
see Zaritsky 1994). If we exclude Leo I from the sample 
(one could claim that it is an
unbound satellite) the satellite that implies the next largest
lower mass limit is Pal 14, $M_{MW} > 4.3 \times 10^{11}$. The
difference
between this limit and that from Leo I is  
substantial, which might lead one to conclude that
Leo I is indeed likely to be unbound. However, Pal 14 is nearly three
times closer to the Galactic center than Leo I, 
so the two satellites are not sampling the
same region of the halo. Because of these ambiguities, and the
fact that the data from all of the other satellites are not being
exploited, this approach is unsatisfactory.

An improvement over the simple binding mass
argument is provided by the projected mass calculation (cf.
Bahcall \& Tremaine 1981). In this approach one combines the
observables
(radial velocity, $v_r$, and distance, $r$)
into a ``mass-like'' variable, $v_r^2r$, and uses either an 
analytic treatment or simulations to recover the expected value of 
this variable. From those 
calculations one can derive the ``correction'' factor necessary to
convert the observed $v_r^2r$ into an unbiased estimate of the
mass.  This approach has several advantages. First, it is
mathematically stable (unlike a virial mass analysis). Second, the
correction factor is easily calculable for various models.
Third, it uses all of the data and can provide a confidence interval
on the final mass rather than just a lower limit. However, this 
approach also
suffers from some of the same problems as methods
discussed previously. For example, it assumes that the test
particles are on bound, relaxed orbits.
The application of a mathematically sophisticated version of
this technique (Little \& Tremaine 1989), using the quantity 
$v^2r$ as the diagnostic variable for a
point mass model, and assuming isotropic orbits, 
results in mass estimates that range from 
$4.6 \times 10^{11} M_\odot$ to $12.5 \times 10^{11} M_\odot$, with
and without the inclusion of Leo I in the sample 
(Zaritsky \etal\ 1989). The large
difference arising from the inclusion or exclusion of Leo I 
has led some to suggest that Leo I is unbound and should not be
included. However, below we demonstrate
that the nature of Leo I has little influence on the final result
if a more realistic halo model is adopted.

Both previous mass estimates (with or without Leo I) 
imply a Galactic halo that extends many tens of kpc. For
an isothermal halo with $v_c = 220$ km sec$^{-1}$, the inferred
extents are 43 and 112 kpc, respectively. In either case, many of the
satellite orbits would penetrate the mass distribution and the
assumption of a point mass potential
is likely to be violated. If we examine
the implications from using an isothermal sphere model, rather than the
point mass model, we arrive at a different conclusion regarding
the influence of Leo I. In the
isothermal
sphere model, one derives $v_c$ and the inclusion or exclusion 
of Leo I makes only a modest difference (154 vs. 169 km sec$^{-1}$, 
respectively, for assumed isotropic orbits). The difficulty with
this model is that it makes no prediction about the extent of the
halo, which formally leads to an infinite halo mass.
Nevertheless, from the best fit isothermal sphere model we calculate
that the enclosed Galactic mass within 200 kpc is $1.3 \times 10^{12}
M_\odot$.

One potential difficulty for any of these models is that they do not account
for the fact that distant satellites, like Leo I,  have only completed
1 or 2 orbits. An assumption that is used in deriving the 
necessary correction
factors is that satellites are distributed randomly
in orbital phase. 
This assumption breaks down when the satellite has only
completed 
a limited number of orbits. The solution to
this problem is to construct models that follow the orbits of 
individual satellites. These models rely on an assumption about the age
of the universe.

\subsection{Timing Arguments}

The timing argument was first applied by Kahn and Woltjer (1959)
to the Milky Way-M 31 system. Because M 31 is moving
toward the Galaxy, one can presume that the gravitational attraction
between M 31 and the Galaxy is sufficient to decouple the pair
from the Hubble expansion. By adopting an age of the Universe
and measuring both the distance to M 31 and its velocity, simple
orbital equations provide a firm lower limit the mass of the 
pair of galaxies. The only potential loopholes in this argument involve
invoking random peculiar velocities for the galaxies or interactions
with other galaxies. 

The application of the simple model (radial orbits, point masses)
results in mass estimates of between 3 and $4 \times 10^{12} M_\odot$
for the M 31-Milky Way pair. Assuming either that the relative
masses of the two galaxies scale with their luminosities or with
their circular velocities squared results in a mass ratio of between
1.3 and 1.7 between M 31 and the Galaxy. Adopting 1.5 as the mass
ratio, the mass of the Milky Way is then inferred to be $\sim 1.4
\times 10^{12} M_\odot$. However, this analysis 
excludes considerations of
angular momentum, the overlap of the two halos at earlier times, and the
growth of the halos with age. All of these considerations will
increase the total masses.

We can also apply the timing argument to the Leo I-MW system. 
In contrast to M 31, which
is falling toward the Galaxy and so one can assume that the system
is seen ``shortly'' after turnaround, Leo I is moving outward and
so one can either assume that it is on its first outward trajectory
(although a simple calculation suggests that it would have traveled
much farther than its current location in a Hubble time) or on its
second outward trajectory. Adopting the second option leads one
to infer a mass for the Galaxy of between 1.1 and $1.5 \times
10^{12} M_\odot$. Again we have neglected angular momentum,
overlapping mass distributions, and the evolution of the Galactic
halo. 

Finally, more complex models attempt to embed the two body system
into the larger environment
(cf. Einasto \& Lynden-Bell 1982, Raychaudhury \& Lynden-Bell 1989,
Peebles 1995, Shaya, Peebles, \& Tully 1995).
By doing this, one can investigate the origin and expected
magnitude of the angular momentum and utilize the dynamics of
other Local Group galaxies to constrain the mass. The results
are all consistent and, as expected, the derived masses are somewhat
larger than those from the simple timing argument. Einasto
\& Lynden-Bell derive $M_{MW} = 1.9\times 10^{12} M_\odot$; 
Raychaudhury and Lynden-Bell derive $M_{MW} = 1.3 \times 10^{12}$;
Peebles (1995)
derives $M_{MW} \sim 2 \times 10^{12} M_\odot$; and Shaya \etal\ derive
$M/L = 175$, which roughly converts to $2.3 \times 10^{12} M_\odot$. 
{\bf The statistical analysis of the satellite galaxies (following
Little \& Tremaine 1989), the MW-M31 timing
argument, the Leo I-MW timing argument, and the analysis of galaxies
out to 3000 km sec$^{-1}$ recessional velocity (Shaya \etal), 
all imply that the
Galactic mass out to $\sim 200$ kpc is $\gtsim 1.2 \times 10^{12}
M_\odot$. }

\subsection{Putting It All Together}

The proper way to constrain the Galactic mass profile
is not to divide the data and
the methods into a myriad of possibilities, but instead to 
treat them all in a single, self-consistent manner. This approach is best
illustrated in the work of Kochanek (1996) who fit
one model to all of the data available at that time. To accomplish
this, he selected the Jaffe model and used a Bayesian statistical
approach following that of Little and Tremaine (1989) to derive
the characteristic scale and circular velocity 
of the Jaffe model, $r_j$ and $v_j$. 
Using all of the data discussed previously, he derives that 
$v_c = 219 [188,251]$ km sec$^{-1}$ and $r_j = 204 [116,359]$ kpc,
where the values in brackets indicate the 90\% confidence limits.
The best fit values imply a mass at 200 kpc of $1.1 \times 10^{12} M_\odot$.
Taking the 90\% confidence limits on $r_j$ and $v_c$ independently, 
we calculate
limits on the mass at 200 kpc of 0.6 and 17.8 $\times 10^{12}
M_\odot$. These are almost certainly conservative values for the 90\% 
confidence limit because it is not evident that good fits (acceptable
within
the 90\% confidence limit)
are obtained when both $v_c$ and $r_j$ are selected
at their individual 90\% confidence limits. As previously described for the
isothermal sphere model, the exclusion of
Leo I makes little difference to the derived parameters.
The results without Leo I
are $v_c = 221 [190,254]$ km sec$^{-1}$ 
and $r_j = 168 [78,321]$ kpc, which results in
an enclosed mass at 200 kpc of $9.8 \times 10^{11} M_\odot$, a
change of $\sim 10$\% in comparison to the enclosed mass derived
if Leo I is included. Kochanek's treatment also demonstrates that
the results are robust (consistent within the uncertainties) to 
excluding or including various sets of data.

\section{Other Galaxies}

We can further test the results for the mass and extent of our
Galaxy by comparing to the results obtained for other galaxies.
The analysis of other galaxies presents its own problems
but circumvents most of those present in the approaches
discussed so far. Zaritsky
\etal\ (1993,1998) and Zaritsky \& White (1994) compiled and analyzed
the dynamics of satellite galaxies for primary galaxies similar
to the Milky Way. In general, the primaries are somewhat more
luminous than the Milky Way and they are Sb to Sc type spirals. 
The satellite sample contains 115 satellites around 69 spiral primaries. The
projected separations and radial velocity differences are plotted
in Figure 2. The philosophy behind the approach is to assume that
the primaries are sufficiently similar that the satellites can
be treated as orbiting a single, typical galaxy. Therefore, even
though there are only one or two satellites per galaxy, 
the analysis of the entire sample has significantly better 
statistics than the analysis of 
of the Galactic  satellite sample.

\begin{figure}
\plotone{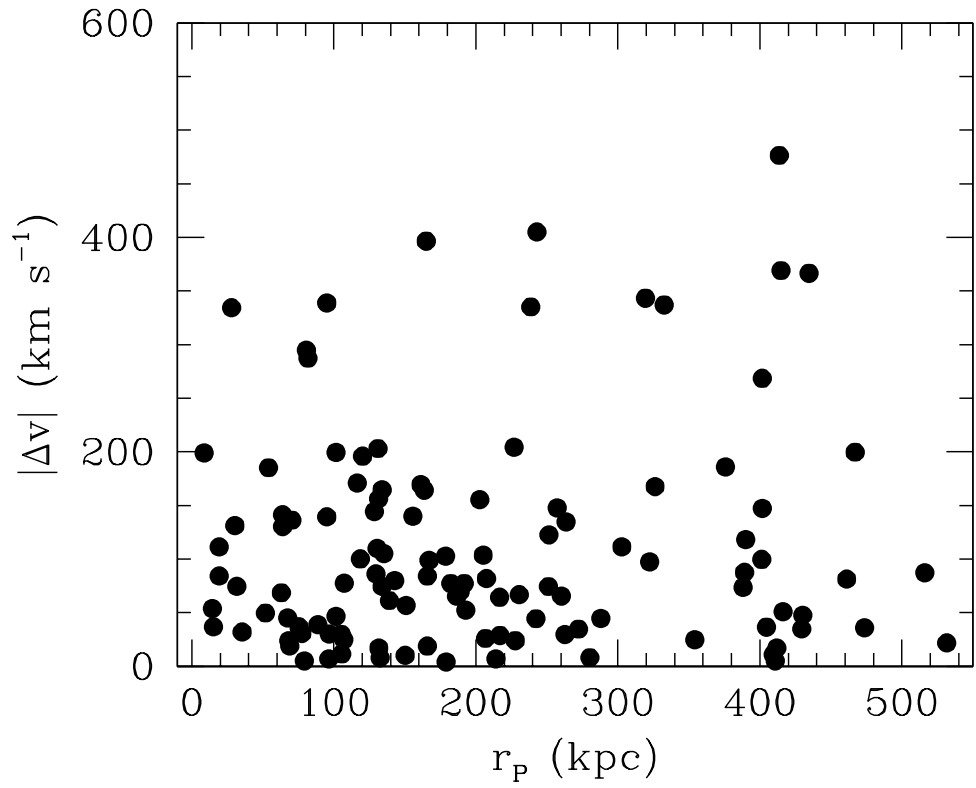}
\caption{The projected separation and absolute value of the
radial velocity differences
for the Zaritsky \etal\ (1998) sample of satellite galaxies.}
\end{figure}

The detailed approach utilized to analyze this sample is based
on modeling the halo growth over time, from its beginning as a seed
perturbation in a uniform background to the current time, 
and the characterization of the properties of test particles 
within the final halo. Models 
that produce a satellite distribution similar to the observed
distribution are accepted. This analysis, conducted on the first sample
of 69 satellites, led to an estimate of the enclosed mass at 
200 kpc of 1.8[1.5,2.3] $\times 10^{12} M_\odot$. The newer
sample is qualitatively similar to the original one
and samples the large radii
better, so the same analysis applied to that sample should produce
similar results. The estimated mass is somewhat larger than the
results quoted above for the Galactic mass enclosed within 200 kpc,
but the primary galaxies in this sample have an average circular
velocity of 250 km sec$^{-1}$. If we presume that halo profiles scale 
as $v_c^2$, then the implied enclosed mass for the Milky Way within 200 kpc
is $1.4 \times 10^{12} M_\odot$, which is perfectly
consistent with Kochanek's results.

\section{Discussion}

Individually, each approach that we have described
has potentially catastrophic
problems. However, these problems and the data
are different in each approach;
therefore, the consistency between various approaches more than 
validates any single approach.
In Figure 3 we plot the results discussed here for comparison.
Each result is plotted at the relevant radius.
Error bars generally represent $\sim$ 90\% confidence
intervals. For the rotation curve datum (Fich and Tremaine) we chose
a range of 180 to 240 km sec$^{-1}$ as representative of the 90\%
confidence interval at $R = 16$ kpc. We plot the quoted 90\% confidence
intervals for Kochanek's measurement of the mass enclosed at 50 kpc. 
For his estimate of the mass enclosed at 200 kpc, we have adopted his
confidence 90\% intervals on the quantities $v_c$ and
$r_j$. Therefore, the plotted uncertainty is probably overestimated because it
is the combination of both variables being at the extreme of their
90\% confidence range. For the measurement of the mass
enclosed at 100 kpc from the analysis of the orbit of the Large
Magellanic Cloud (Lin, Jones, \& Klemola) we adopt their error 
estimate (for which the confidence level is unclear). Two values
are overplotted from Zaritsky \& White (ZW) for $H_0 =75$ km
sec$^{-1}$. 
The upper one is the value
of the mass derived for the average galaxy in the sample. The lower
has been corrected by the square of the disk circular velocity of the
average ZW galaxy (250 km sec$^{-1}$) relative to the Milky 
Way. ZW concluded that halos are not correlated with disk rotation
speeds (so it is unknown whether the correction for the circular
velocity difference is appropriate). Three values are plotted from
the Zaritsky \etal\ (ZOSPA) study (the one at slightly
smaller radius is derived assuming 
radial orbits, the central one is the lower limit derived by applying
the timing argument to Leo I, and the one plotted at the slightly
larger radius
is derived assuming isotropic orbits). The limits on the Einasto \& 
Lynden-Bell results were taken from their preferred range of solutions
with the additional limitation that the age of the universe is between
10 and 15 Gyr. Peebles's paper did not quote uncertainties. For the 
Shaya \etal\ result we adopted the 2$\sigma$ contour from their model
with external mass perturbers and draw an arrow to the right
indicating that this value applies to large unspecified enclosed radii. 
{\bf All of the data in Figure 3 are entirely consistent with
an isothermal sphere with $v_c \sim 180$ km sec$^{-1}$.
There is no evidence for a significant truncation of the mass
profile at large radii.}

\begin{figure}
\plotone{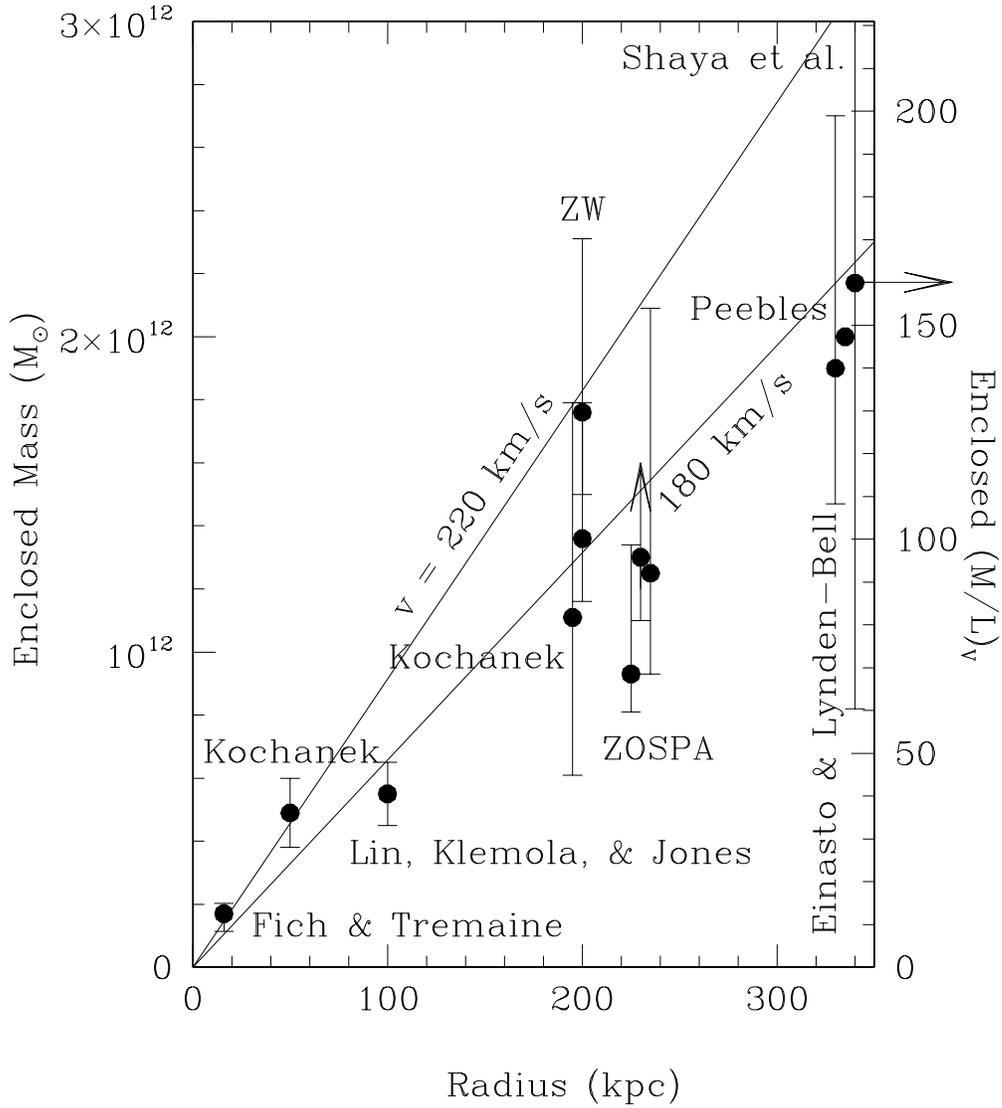}
\caption{A comparison of various measurements of the mass of the
Galactic halo (see text for details). The solid lines illustrate the expected enclosed mass for isothermal
spheres with $v_c = 180$ and 220 km sec$^{-1}$. ZW and ZOSPA stand for
Zaritsky \& White (1994) and Zaritsky \etal\ (1989), respectively.}
\end{figure}

Our Galaxy has a M/L ratio (for 
the mass enclosed within 200 kpc) of $\sim$ 100 in solar
units. Because of the presence of M 31, the Galactic halo cannot
extend much beyond 200 kpc and $M/L$ is unlikely to be more than
200. If all galaxies have $M/L \sim 100$, 
then $\Omega_{galaxies} \sim 0.07$, which is well below the critical
value but well above the limit on $\Omega_{baryons}$ of 0.0193
(Burles and Tytler 1998). 
If we accept the conclusion from the MACHO experiment 
(Alcock \etal\ 1997) that
50\% of the Galactic dark matter is baryonic, this predicts that
$\Omega_{baryon} \sim 0.035$ which is significantly larger than
the limit from the deuterium observations.
Four possible solutions to this problem are (1) the halo
baryonic component is highly concentrated toward the center of
the Galactic halo (so that it is $\ll $50\% of the {\it total} dark
matter in the halo but $\sim$ 50\% of the dark matter to the radius
of the Large Magellanic Cloud), 
(2) the dark matter is composed of primordial black holes (which
are baryonic but which do not participate in big bang
nucleosynthesis),
(3) the halo dark matter is highly clustered (Widrow \& Dubinski 1998)
and the line-of-sight
to the Large Magellanic Cloud intersects a high density feature (which
would again imply that the baryonic matter is $\ll$ 50\% of the total halo),
or (4) the interpretation
of the microlensing events as originating from halo MACHOs is 
incorrect (Sahu 1994, Zhao 1998, Zaritsky \& Lin 1998).

\section{Conclusions}

All of the measurements of the mass of the Galactic halo are 
consistent if one accounts for the fact that they reflect results
from  different
radial ranges within the halo. A simple isothermal sphere model
fits the data from 10 kpc to 300 kpc. 
An ``isothermal-like'' model where the characteristics
rotation curve drops slightly ($\sim$ 20\%) from the standard disk
value 220 km sec$^{-1}$ at radii $\gtsim$ 20 kpc is entirely 
consistent with all of the data.
Barring fundamental problems
with our understanding of gravity, the Galactic halo extends 
at least $\sim$ 200 kpc and contains $\gtsim 10^{12} M_\odot$.
Models that extrapolate the declining
rotation curve between 10 and 20 kpc 
outward, and so predict a low halo mass, are 
incompatible with the orbit of the Magellanic Clouds, with the
dynamics of the distant Galactic satellites, with the dynamics of
the Local Group, and with the halo properties inferred for
other spiral galaxies. Despite the mystery regarding the nature
of the dark matter, the measurement of its distribution around 
spiral galaxies is now secure.

\acknowledgments

DZ thanks his collaborators on  various aspects of his work
discussed here (Marc Aaronson, Carlos Frenk, Ed Olszewski, Ruth
Peterson, Bob Schommer, Rodney Smith, and Simon White). 
DZ acknowledges financial support from NASA through HF-1027.01-91A
from STScI, which is operated by AURA, Inc. under NASA contract
NAS 5-26555, the David and Lucile Packard Foundation, 
the Sloan Foundation, and the conference organizers.

%Check recent ASP Conference Series for appropriate bibliography structure.
%paspconf.doc and paspconf.sty list useful shortforms for all the
%major journals.

\end{document}